\documentclass[a4paper,11pt]{article}
\usepackage{pos}
\usepackage[capitalize]{cleveref}
\usepackage{xspace}
\usepackage{multicol}
\usepackage{enumitem}
\usepackage{wrapfig}

\newcommand{\pulsestart}{\ensuremath{t_\text{pulse}^\text{start}}\xspace}
\def\Offline{\mbox{$\overline{\textrm{Off}}$\hspace{.05em}\protect\raisebox{.4ex}{$\protect\underline{\textrm{line}}$}}\xspace}

\title{Characterizing the neutron component of extensive air showers with the Surface-Scintillator Detectors of AugerPrime}
\ShortTitle{Neutron measurement in EAS}

\author*[a]{Tobias Schulz}
\onbehalf{for the Pierre Auger Collaboration${}^b$} 

\affiliation[a]{Institute of Physics of the Czech Academy of Sciences, Na Slovance 1999/2, 182 00 Prague, Czech Republic}
\affiliation[b]{Observatorio Pierre Auger, Av.\ San Mart{\'\i}n Norte 304, 5613 Malarg\"ue, Argentina\\
Full author list: {\rm\url{https://www.auger.org/archive/authors_icrc_2025.html}}}
\emailAdd{spokespersons@auger.org}

\abstract{Neutrons are the only neutral hadrons that remain stable over the timescale of an air-shower development.
Their energy is lost only through hadronic interactions and quasi-elastic scattering, which results in their high abundance at the ground.
The signals from the electromagnetic and muonic components in scintillation detectors typically span only a few microseconds.
In contrast, the neutrons can cause delayed pulses in scintillation detectors up to and beyond several milliseconds after the passage of the shower front.
Selection of an appropriate time window allows us to isolate and characterize the neutron component of air showers, which may provide a new, direct method to probe hadronic interactions during the shower development.
We report the measurement of a neutron component at ultra-high energies using the Surface-Scintillator Detectors (SSD) from the AugerPrime upgrade of the Pierre Auger Observatory.
We provide a first look at the pulse-amplitude spectrum together with our measured rate and lateral distribution of the neutron component.}

\ConferenceLogo{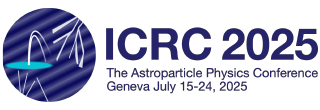}

\FullConference{39th International Cosmic Ray Conference (ICRC2025)\\
 15–24 July 2025\\
Geneva, Switzerland\\}

\begin{document}
\maketitle

\section{Introduction}

The measurement of extensive air showers and the separation of their components are important tools for testing hadronic interaction models at ultra-high energies.
The models must be extrapolated for the simulation of air showers because the primary cosmic rays can exceed the energies achievable by human-made accelerators by several orders of magnitude.
An air shower can be divided into electromagnetic and hadronic shower components. 
From the hadronic shower component, neutrons are the only relevant massive and neutral particles that are stable long enough to reach the ground.
Simulations of extensive air showers have shown that neutrons with a broad energy spectrum can reach the ground with possible delays of up to several milliseconds following the initial shower front~\cite{Erlykin:2006, Schimassek:2023}.
Pulses with such large delays were measured and documented in 1984 by John Linsley with the scintillator detectors of the Volcano Ranch experiment~\cite{Linsley:1984}.
He assumed these pulses could correspond to neutrons traveling at speeds slightly below the speed of light, an idea previously suggested by Kenneth Greisen~\cite{Greisen:1962}.
Since neutrons are neutral particles, they can produce a signal in scintillator only through elastic and inelastic scattering with nuclei in the scintillator material or via neutron-capture processes that emit gamma rays.
In the upgrade of the Pierre Auger Observatory~\cite{ThePierreAuger:2015} known as \emph{AugerPrime}~\cite{Aab:2016}, \emph{Surface-Scintillator Detectors} (SSDs) were installed atop the existing \emph{Water-Cherenkov detectors} (WCDs).
Furthermore, new electronics for data-acquisition, featuring an increased sampling rate of 120\,MHz compared to the original rate of 40\,MHz, were installed.  
During early SSD measurements with the pre-production array, delayed pulses were observed in the time traces~\cite{PAO:2021}.

\section{Data collection and candidate selection}

The optimal efficiency of the SSD in detecting neutron pulses is expected to begin at energies around a few MeV.
Below 0.7\,MeV, the probability of detecting a neutron in the scintillator drops below 0.001\%. 
Conversely, at energies above 100\,MeV, the efficiency in detecting neutrons reaches approximately 2.3\%~\cite{Schimassek:2023}.
The available data from the AugerPrime upgrade are used for this analysis, comprising measurements from 2020 through March 2025.
The minimal reconstructed shower energy is set to $10^{18}\,$eV and the zenith angles of the showers are limited to a maximum of $60^\circ$.
This selection yields a dataset of 262\,497 events with 837\,543 time traces.
Only traces from the high-gain channel of the SSD photomultiplier tube (PMT), which are free from baseline saturation caused by undershoot, are included in the analysis.
The peak and integral of a signal produced by a vertical minimal ionizing particle are defined as $\text{MIP}_\text{peak}$ and $\text{MIP}_\text{charge}$, respectively.
A simple pulse-finding algorithm is applied to each trace.
A typical pulse charge from a single photoelectron is approximately $0.05\,\text{MIP}_\text{charge}$.
The pulse start time \pulsestart is defined as the moment when a bin value in the trace exceeds a conservative value of $0.8\,\text{MIP}_\text{peak}$.
The pulse end time is determined when the bin value drops below $0.8\,\text{MIP}_\text{peak}$ for two consecutive bins to mitigate the effects of afterpulsing or electronic noise.
The pulse start and end times are measured relative to the signal start time $t_\text{start}$, determined by the WCD.
The charge $S_\text{pulse}$ for each pulse is obtained by integrating the signal from the start to the stop bin, expressed in units of $\text{MIP}_\text{charge}$.
If a pulse is sufficiently large to saturate the trace, the low-gain channel is used instead to estimate $S_\text{pulse}$.

\begin{figure}
\centering
\includegraphics[width=1\textwidth]{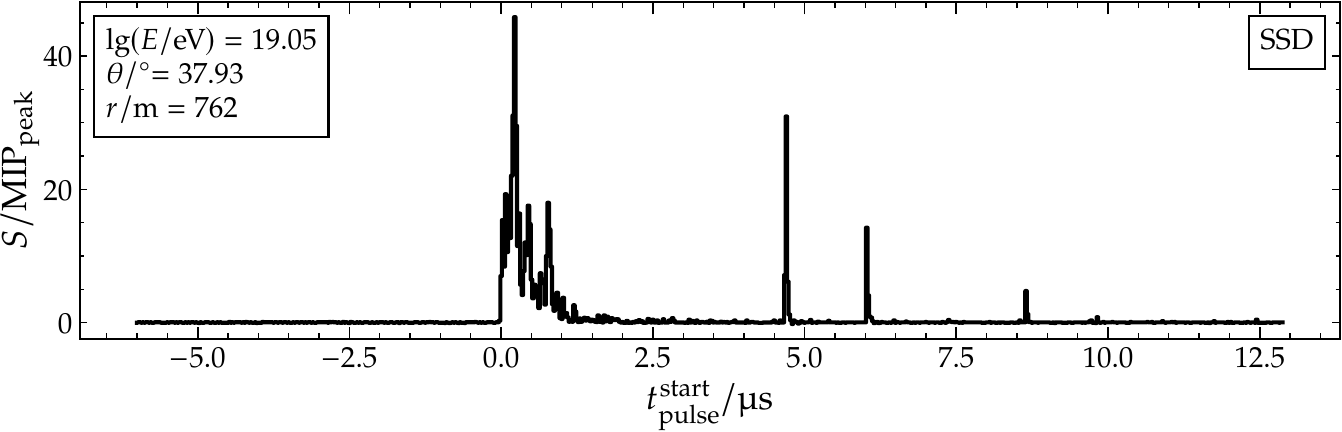}
\caption{An example SSD trace is shown for a station participating in an event of the pre-production array. 
The time \pulsestart is adjusted to commence at the estimated start of the passage of the shower front.
Three large, late pulses are notable in the SSD trace.}
\label{fig:PulseDataSet}
\end{figure}

In \cref{fig:PulseDataSet} an example time trace of the SSD with visible late pulses is shown.
At $\pulsestart = 0$, the passage of the shower front generates a broad signal.
Pulses that arrive before this would likely be caused by atmospheric muons.
The majority of pulses occur within approximately $2.5\,\upmu$s. 
These late pulses can reach charges up to $220\,\text{MIP}_\text{charge}$ with an average duration of around $0.03\,\upmu$s.
A time threshold of $\pulsestart \geq 5\,\upmu$s is set as well as a minimum pulse duration of $16.\bar{6}$\,ns to select candidate pulses that may originate from delayed neutrons.
This threshold is chosen to be sufficiently long after the initial shower front to resolve individual pulses.

Additionally, the background contribution from other particles of electromagnetic or muonic origin is quantified using air shower simulations.
60\,000 showers are simulated, using the \textsc{Corsika} simulation software~\cite{Heck:1998}.
The events consist of a mixture of 4 primary particles (proton, helium, oxygen, iron) in a 1:1:1:1 ratio.
They are distributed uniformly over $\sin^2\theta$, with $\theta$ ranging from 0 to $60^\circ$, and energy, ranging from $10^{18}\,$eV to $10^{19.5}\,$eV.
The utilized hadronic interaction model is \textsc{Epos-LHC}, and the hadron energy cut-off in the simulated showers is set at 20\,MeV.
The \Offline software framework~\cite{Offline} is utilized for the detector simulation and shower reconstruction.
To increase the number of recorded traces in the simulations, additional stations are simulated in concentric ``dense rings'' around the shower axis.
Eleven such dense rings are generated, each comprising 4 stations, and are logarithmically spaced between 100 and 1700\,m from the shower axis.
In the distance range of 400 to 800\,m from the shower axis, for pulses exceeding approximately $11\,\text{MIP}_\text{charge}$, the background of non-hadronic pulses is around 3\%.
This background increases to about 15\% for pulses below $11\,\text{MIP}_\text{charge}$.
Depending on the shower energy, zenith angle, distance to the shower axis, and the charge of the individual pulses, the fraction of pulses originating from muonic and electromagnetic particles after $5\,\upmu$s ranges from approximately 1 to 30\%.
Although neutrons outnumber protons on the ground by a factor of 100, the probability that a neutron produces a detectable signal in a scintillator is at most around 17\% depending on its energy~\cite{Schimassek:2023}.
However, the vast majority of these protons originate from neutron decay.

\section{Measurement of subluminal pulse rate}

\begin{figure}
\centering
\def\h{0.38}
\includegraphics[height=0.39\textwidth]{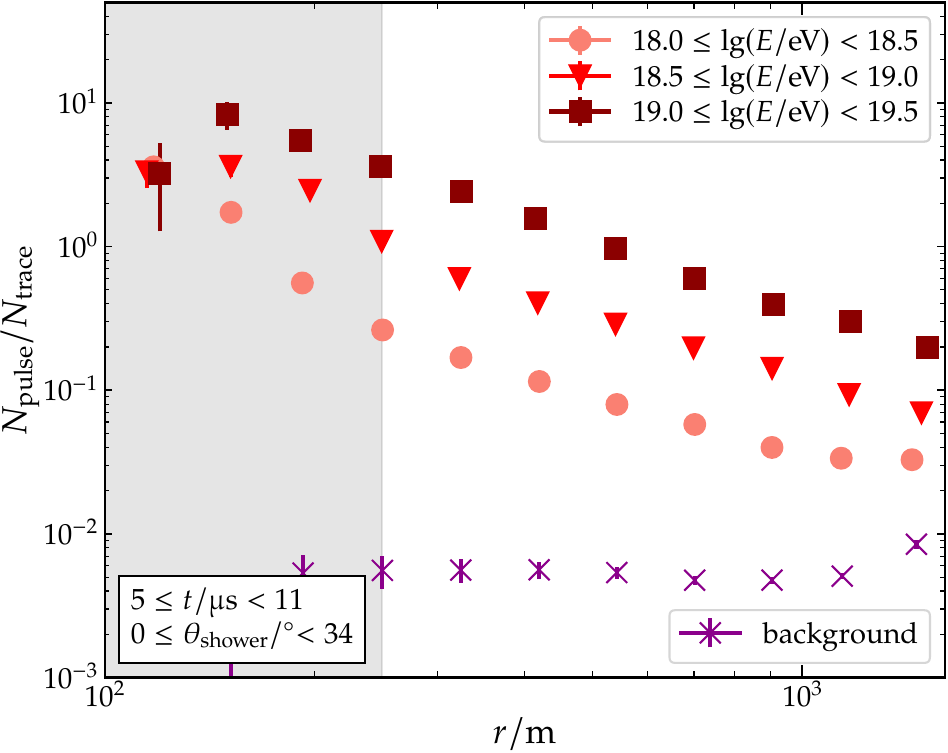}\hfill
\includegraphics[height=0.383\textwidth]{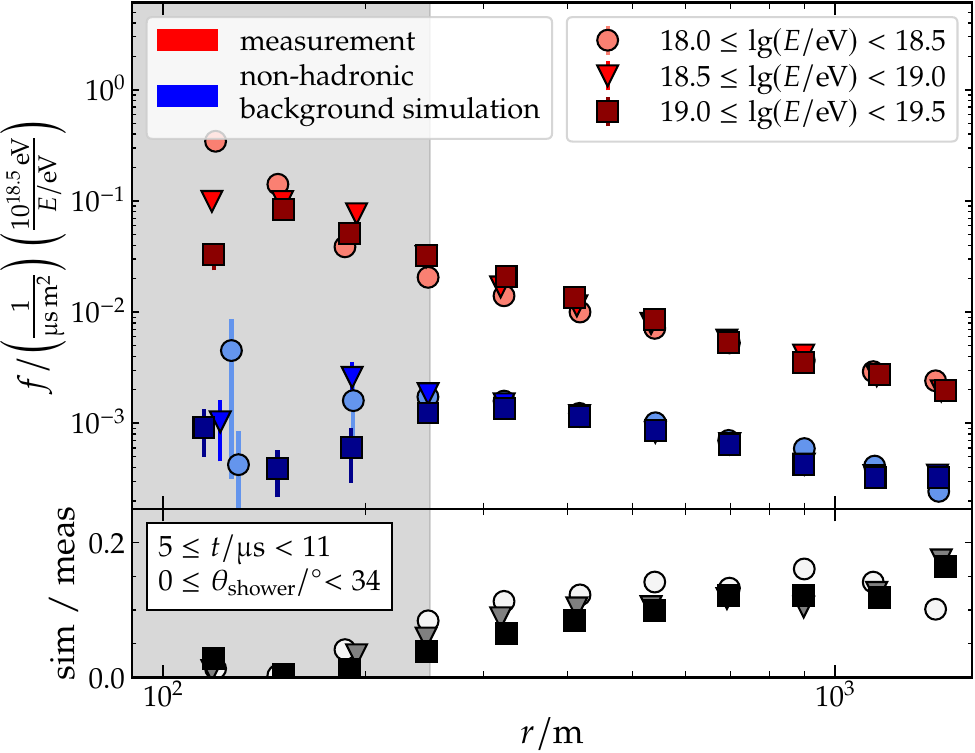}
\caption{\emph{Left:} The average rate of candidate pulses $N_\text{pulses}$ per trace increases with energy and decreases with distance from the shower axis.
Below around 250\,m from the shower axis (shaded area), the electronics baseline can saturate due to undershoot and pulses can not be counted correctly.
\emph{Right:} The pulse rate is normalized to the observation time window, the scintillator area and the shower energy.
From the simulated showers, the normalized pulse rate $f$ of non-hadronic particles is expected to contribute between 5 and 18\% as background to the measurements.}
\label{fig:NeutronPulseRates}
\end{figure}

In \cref{fig:NeutronPulseRates}-left, the average number of candidate pulses $N_\text{pulses}$ per trace as a function of distance from the shower axis is shown for events between zenith angles of 0 and 34$^\circ$.
In the energy range from $10^{18.5}\,$eV to $10^{19}\,$eV and at a distance of $\approx$1000\,m from the shower axis, a total of 743 late pulses are detected across 6129 traces, resulting in an expected pulse rate of approximately 12\% per trace.
At a closer distance of around 600\,m, this rate increases to 24\% with 574 late pulses observed in 2393 traces.
In the higher energy range from $10^{19}\,$eV to $10^{19.5}\,$eV at around 1000\,m from the shower axis, 278 late pulses are recorded in a total of 809 traces, yielding a pulse rate of 34\%.
Similarly, at around 600\,m, the pulse rate rises to 88\% with 261 late pulses detected in 297 traces.
All pulses arriving at $\pulsestart < -0.2\,\upmu$s are classified as background pulses unrelated to the shower.
Their rate is calculated and then normalized to the counting window of the late pulses.
At the closest distances to the shower axis, the late pulse rate surpasses the background pulse rate by over two orders of magnitude.
The rate begins to plateau at distances below $250\,\text{m}$ from the shower axis.
This behavior can be attributed to the saturation of the electronics baseline due to undershoot.
The fraction of stations with a saturated baseline $N_\text{sat}/N_\text{tot}$ rapidly increases, starting at approximately 250\,m from the shower axis. 
Since pulses cannot be reliably counted in traces with a saturated baseline caused by undershoot, a reduced counting rate in this region is expected.

As previously mentioned, the pulse rates include pulses from a potential non-hadronic component.
They are contaminated with pulses from muonic or electromagnetic components, which are not accounted for in the background estimation of \cref{fig:NeutronPulseRates}-left.
This additional background varies with the zenith angle, shower energy, and pulse signal.
In \cref{fig:NeutronPulseRates}-right, a normalized pulse rate $f$ is calculated, by accounting for the area of the scintillator, the observation time window and the shower energy. 
Additionally $f$ of non-hadronic particles is calculated from the simulated showers to estimate a background expectation to the measurements. 
In the lower plot of \cref{fig:NeutronPulseRates}-right the ratio between the simulations and measurements are shown. 
For all energies, the expected background contribution of non-hadronic particles to the pulses is between 5 and 18\%.

\section{Measurement of subluminal pulse amplitude spectrum}

\begin{figure}
\centering
\def\h{0.38}
\includegraphics[height=\h\textwidth]{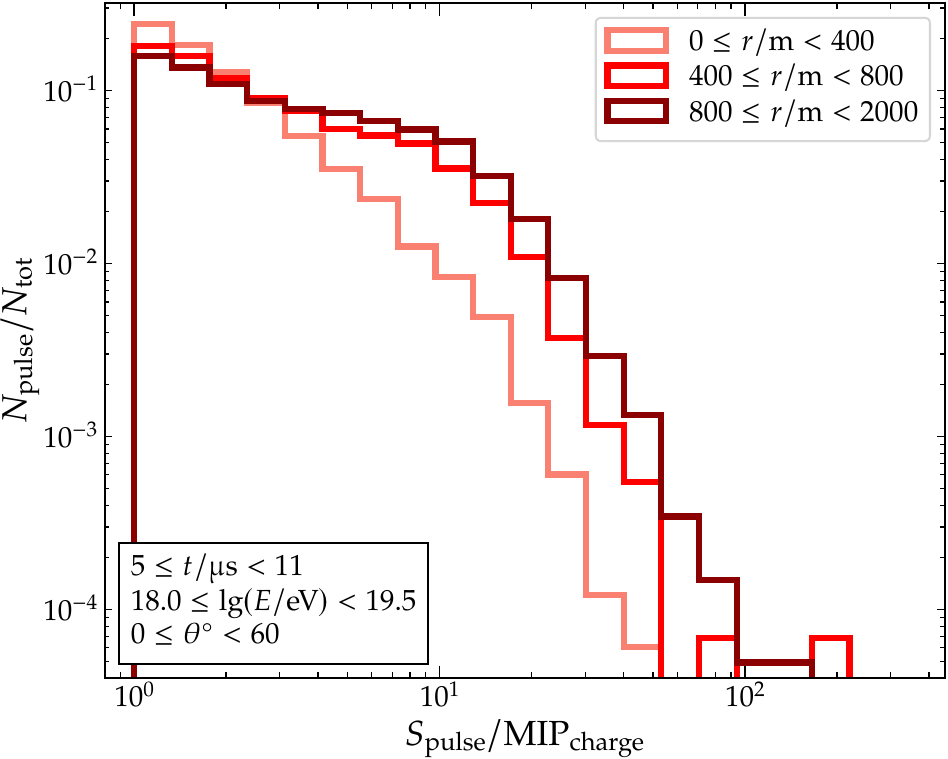}\hfill
\includegraphics[height=\h\textwidth]{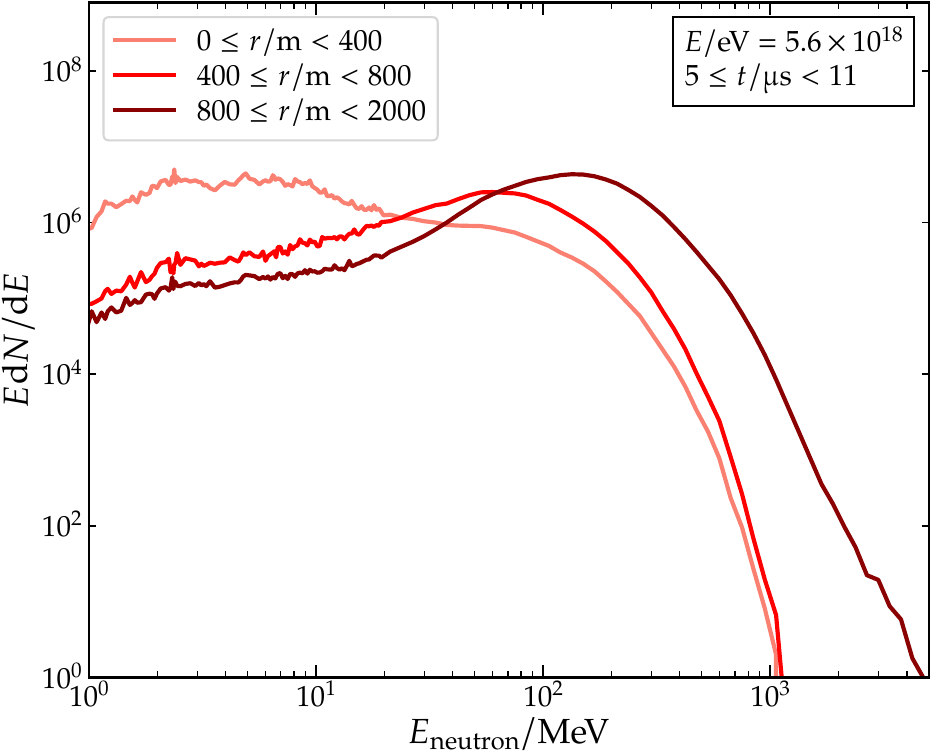}
\caption{\emph{Left:} The measured pulse spectrum for three distance bins. 
With increasing distance to the shower axis, larger $S_\text{pulse}$ can be observed.
\emph{Right:} Three simulated neutron spectra of Ref.~\cite{Schimassek:2023}, with a comparable parameter space as for the measurements.
In the observed time range between 5 and 11\,$\upmu$s, the energy spectrum of neutrons extends to larger energies.}
\label{fig:NeutronPulseSpectrum_Distances}
\end{figure}

The individual charges, $S_\text{pulse}$, of all candidate pulses measured are combined to obtain a pulse amplitude spectrum. 
To compare the shapes of different pulse spectra -- categorized into bins based on energy, shower zenith angle, and distance -- the number of pulses $N_\text{pulse}$ in each bin is normalized to the total number of pulses $N_\text{tot}$ within the corresponding signal bin.
No apparent dependence of the pulse spectrum shape on varying shower energies or zenith angles is observed.
Consequently, the energy and zenith angle bins are combined to improve statistical significance.
In \cref{fig:NeutronPulseSpectrum_Distances}-left, the pulse amplitude spectra for three distance bins, ranging between 0 and 400, 400 and 800, and 800 and 2000\,m from the shower axis are shown.
As the distance from the shower axis increases, the pulse amplitude spectrum extends to larger $S_\text{pulse}$.

In Ref.~\cite{Schimassek:2023}, the \textsc{Fluka} simulation framework~\cite{Ferrari:2005, Bohlen:2014} was used to demonstrate that the shape of the neutron energy spectrum varies depending on both the arrival time relative to the shower signal and the distance from the shower axis.
In \cref{fig:NeutronPulseSpectrum_Distances}-right simulated neutron spectra from Ref.~\cite{Schimassek:2023} at a shower energy of $5.6{\times}10^{18}$\,eV across three distinct distance bins, corresponding to those used in the measurements, are shown.
Within the chosen time window of 5 to 11\,$\upmu$s, the energy spectrum extends to higher energies as the distance from the shower axis increases.

\begin{figure}
\centering
\def\h{0.38}
\includegraphics[height=\h\textwidth]{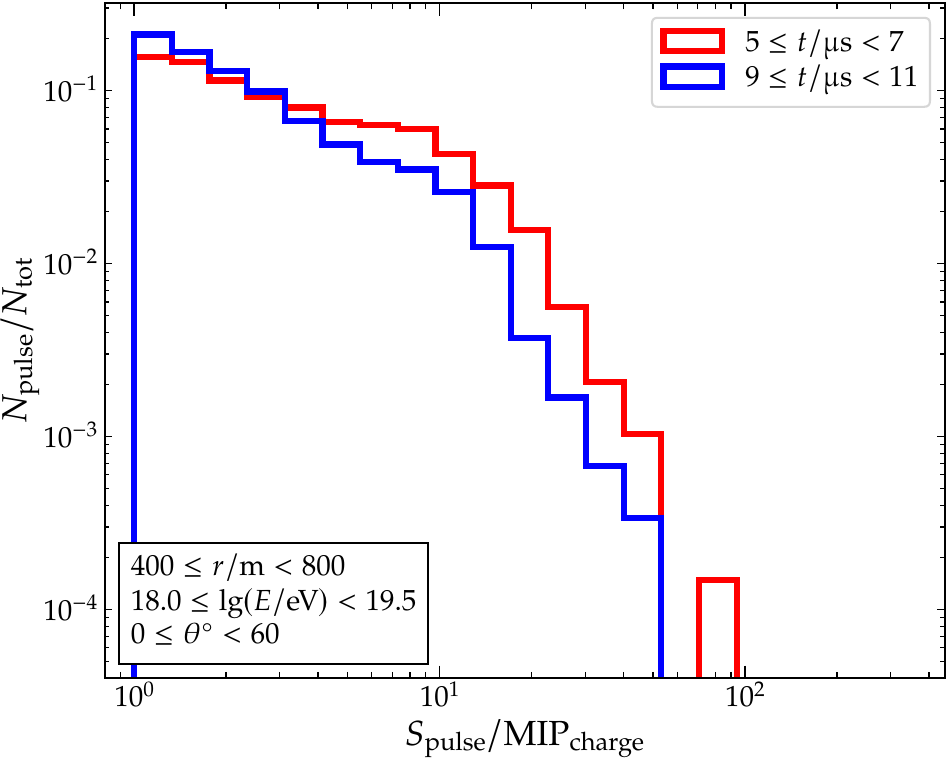}\hfill
\includegraphics[height=\h\textwidth]{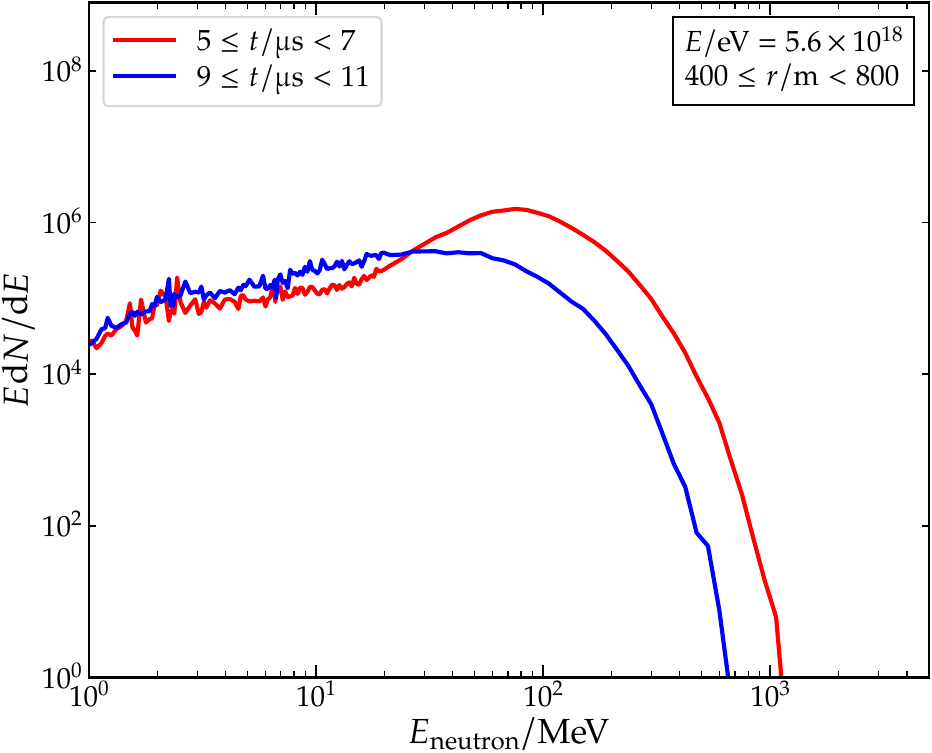}
\caption{\emph{Left:} When binning the pulse spectrum for different pulse arrival times, the spectrum, containing earlier (5 to 7\,$\upmu$s) pulses, reaches to larger $S_\text{pulse}$ than when considering later (9 to 11\,$\upmu$s) pulses.
This trend corresponds to the expectation that neutrons arriving earlier on the ground have larger energies than neutrons arriving later.
\emph{Right:} Two simulated neutron spectra of Ref.~\cite{Schimassek:2023}, with a comparable parameter space as for the measurements.}
\label{fig:NeutronPulseSpectrum_Timing}
\end{figure}

The pulse amplitude spectrum for the distance bin between 400 and 800\,m is now separated between `early' and `late' pulse arrival times.
Early arrival times are defined between 5 and 7\,$\upmu$s, while late arrival times are between 9 and 11\,$\upmu$s.
In \cref{fig:NeutronPulseSpectrum_Timing}-left, the pulse amplitude spectra for these two arrival time bins are displayed, showing a slight hardening of the spectrum for the early arrival times.
In \cref{fig:NeutronPulseSpectrum_Timing}-right, two simulated neutron spectra of Ref.~\cite{Schimassek:2023} are shown for a shower energy of $5.6{\times}10^{18}$\,eV and a distance from 400 to 800\,m to the shower axis and the same arrival time bins as used in the measurements.
The simulated neutron energy spectra exhibit similar behavior to the measured pulse amplitude spectra, which contain more pulses above a few MIP measured at later arrival times.

\section{Towards the forward-folding of neutron energy spectra}

The shape of the pulse amplitude spectrum is strongly dependent on the shape of the input neutron energy spectrum~\cite{Schulz:2025}.
Therefore, a first simulation attempt of possible pulse amplitude spectra from neutrons and protons in the SSD is performed.
To achieve this, single hadrons with fixed energies $E$ are injected into the SSD using the \Offline framework.
The energies range from 1\,MeV to 4700\,MeV and are divided into 54 logarithmic bins.
For each energy bin separate simulations 500\,000 neutrons and 500\,000 protons are performed at a fixed impact angle of $45^\circ$.
Additionally, Birks' law~\cite{Birks:1951} is applied to account for quenching effects in the scintillator.
It describes the non-linear dependence of the scintillation light yield on the energy loss of charged particles in a scintillator.
A Birk's coefficient of ${\sim}0.126$\,mm/MeV for polystyrene-based scintillators~\cite{PDG} is used as a first approximation.
The resulting signal pulses, $S_\text{pulse}$, in the SSD are then analyzed using the same pulse-finding algorithm used for the neutron measurements.
An input energy spectrum from Ref.~\cite{Schimassek:2023} is used, based on the particle distribution of a vertical shower with an energy of $5.6{\times}10^{18}$\,eV at an atmospheric depth of about 878\,g/cm$^2$, matching the depth at which the measurements were taken.
This spectrum includes all downward-going particles located within 400\,m to 800\,m from the shower axis and arriving between 5\,$\upmu$s and 11\,$\upmu$s.
A scaling factor is determined for each input energy spectrum and its corresponding energy bin.
The scaled pulse spectra from these simulations are then combined to form the total pulse spectrum, which is subsequently normalized to the total number of pulses.
This enables a comparison with the measurements. 
However, this comparison cannot be considered exact, as the procedure is limited to a fixed particle impact angle, whereas in reality, a wide distribution of impact angles is expected, depending on particle energy and distance from the shower axis.

\begin{figure}
\centering
\def\h{0.38}
\includegraphics[height=\h\textwidth]{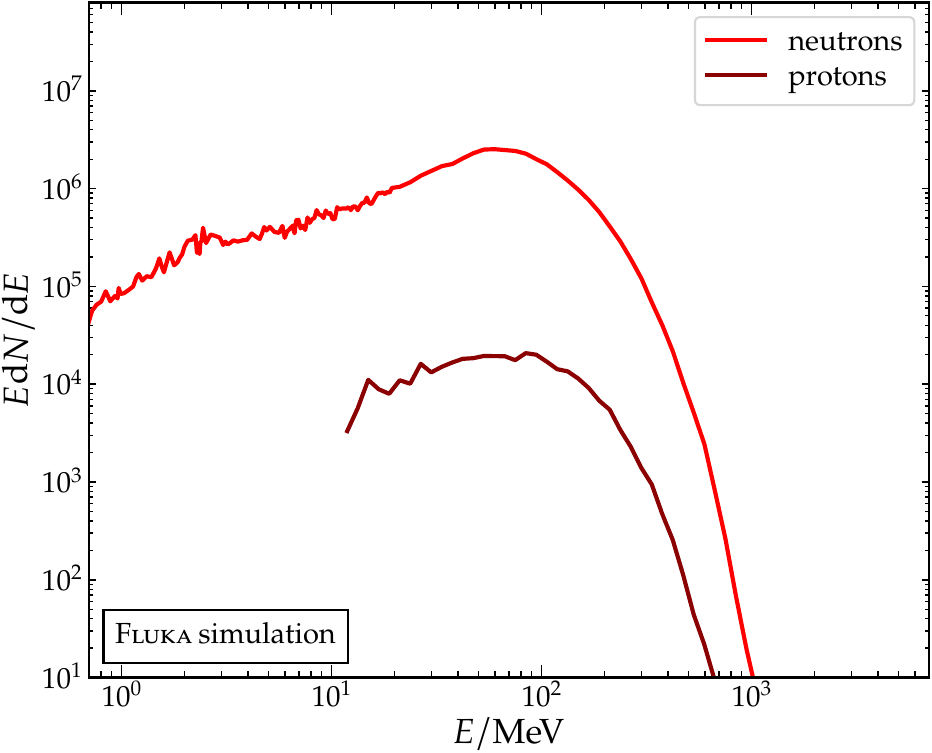}\hfill
\includegraphics[height=\h\textwidth]{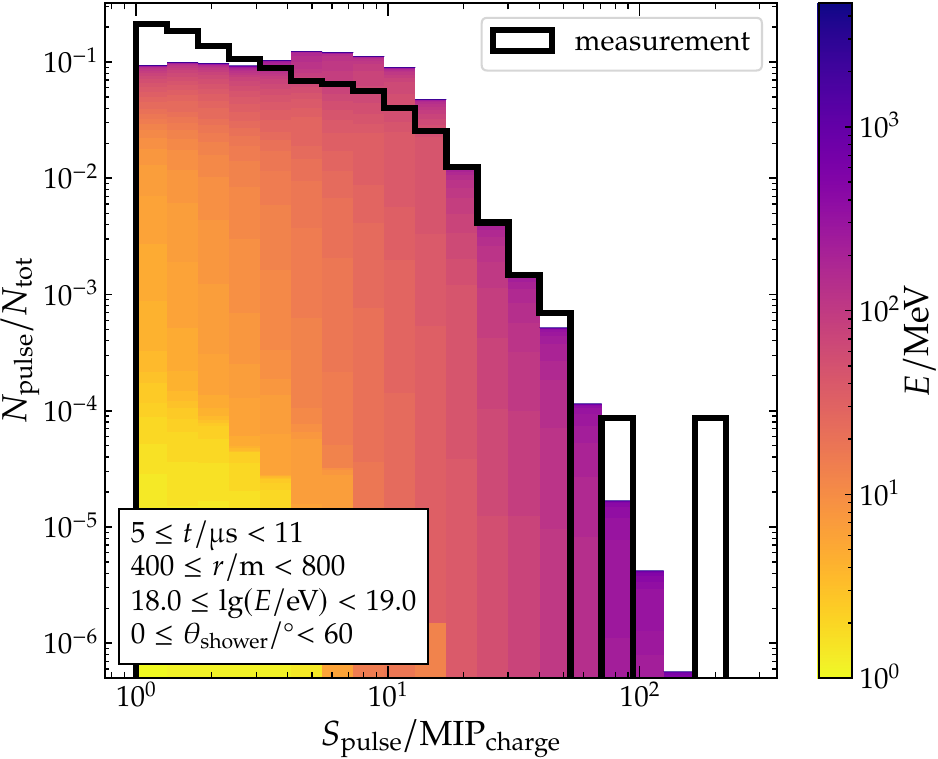}
\caption{\emph{Left:} The simulated pulse amplitudes are sampled from different input energy spectra.
The distribution of protons originates mainly from neutron decay.
\emph{Right:} The resulting pulse amplitude spectra can then be compared with the measured pulse spectrum.
The contribution of each particle energy is color-coded to visualize its relative contribution to the full spectrum.}
\label{fig:NeutronSpectraForwardFold}
\end{figure}

The input energy spectra and the resulting pulse amplitude spectrum are shown in \cref{fig:NeutronSpectraForwardFold}.
In the simulation library, protons with energies below 10\,MeV do not reach the SSD due to ionization losses. 
Therefore, the proton energy spectrum has a lower cutoff at 10\,MeV. 
In the resulting pulse amplitude spectrum the contribution of each particle energy is color-coded to visualize its relative impact on the full spectrum.

\begin{figure}
\centering
\def\h{0.38}
\includegraphics[height=\h\textwidth]{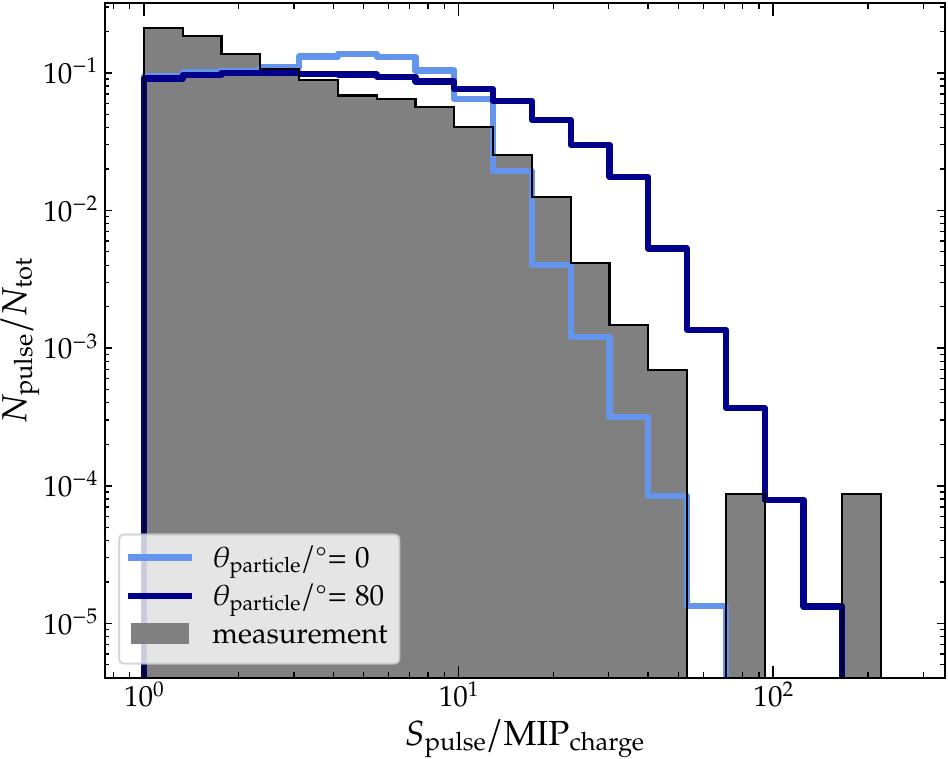}\hfill
\includegraphics[height=\h\textwidth]{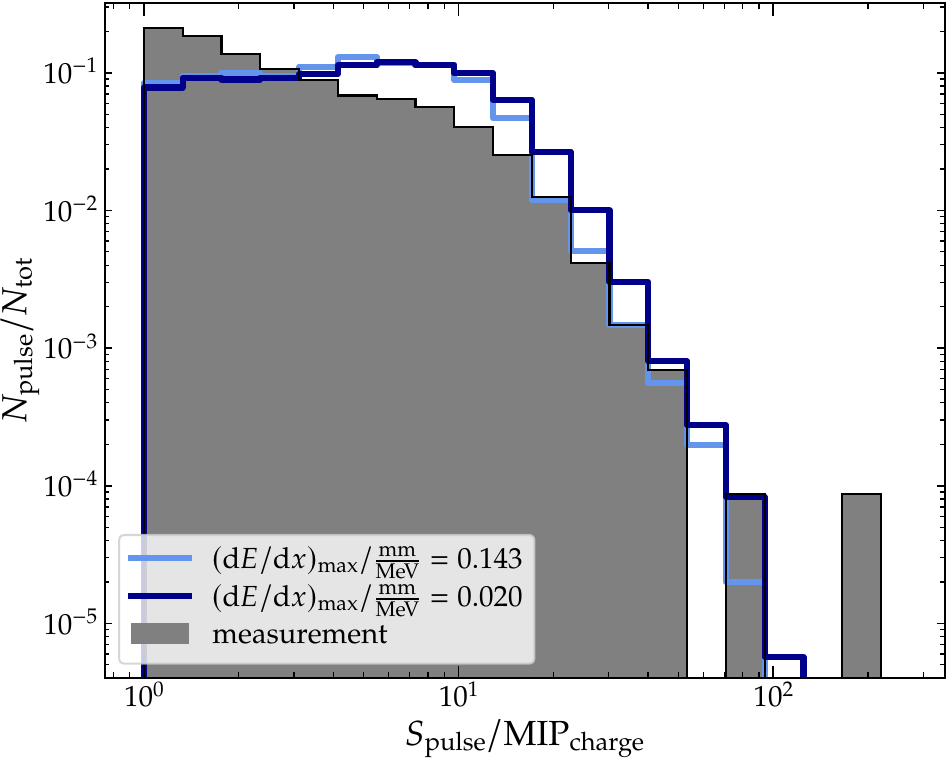}
\caption{\emph{Left:} The shape of the pulse amplitude spectrum depends on the impact angles of the particles.
While a smaller impact angle results in an increase of pulses below approximately $10\,\text{MIP}_\text{charge}$, a larger impact angle will significantly increase the amount of pulses above $10\,\text{MIP}_\text{charge}$.
\emph{Right:} A change in the quenching coefficient has only a minor effect on the shape of the pulse amplitude spectrum.}
\label{fig:PulseAmplitudeAngleQuenchCompare}
\end{figure}

The simulations are repeated with varying particle impact angles and quenching coefficients to examine their influence on the simulated pulse amplitude spectrum.
In \cref{fig:PulseAmplitudeAngleQuenchCompare}-left, a comparison between simulated spectra with fixed impact angles of 0 and $80^\circ$, along with the measured pulse amplitude spectrum is shown.
For amplitudes above approximately $12\,\text{MIP}_\text{charge}$, the measurements lie within the bounds of the simulations. 
As the impact angle increases, the number of pulses with larger signal also increases.
Likewise, simulated spectra for quenching coefficients of 0.143 and 0.02\,mm/MeV are shown in \cref{fig:PulseAmplitudeAngleQuenchCompare}-right.
Compared to the variation caused by the different impact angles, the effect of changing the quenching coefficient is relatively minor.
The forward-folded pulse amplitude spectrum could be further refined by using a more realistic angular distribution of particles rather than relying on a fixed impact angle, and by obtaining a more accurate estimation of the quenching coefficient.

\section{Conclusions}

Late pulses are observed in measurements from the new AugerPrime SSD dataset, revealing rates up to three orders of magnitude higher than the expected background pulses that are not related to the shower.
Simulations of air showers indicate that the fraction of pulses originating from muonic and electromagnetic components ranges from 1\% to 30\%, depending on the primary energy, zenith angle, and distance from the shower axis.
Additionally, the pulse amplitude spectrum of the late pulses was examined, and its shape appears to be largely independent of both primary energy and zenith angle. 
However, with increasing distances from the shower axis the spectrum extends to larger pulse amplitudes.
Using the \Offline simulation framework, pulse amplitude spectra from the SSD signal responses for neutrons and protons at different energies and impact angles were generated.
It is demonstrated how the pulse amplitude spectrum can be modeled by forward-folding energy spectra of protons and neutrons at the ground, as well as the dependence on the particle impact angle and quenching of the SSD.
A comparison between the SSD measurements and \textsc{Fluka} simulations of the particle energy spectrum from Ref.~\cite{Schimassek:2023} showed first-order agreement.
The measurement of late pulses in the SSD of the AugerPrime upgrade suggests that the neutron component of air showers can be further studied at the Pierre Auger Observatory.
While the rate of late pulses is likely too low to permit event-by-event analysis at primary energies below $10^{19.5}$\,eV, the cumulative data from the expanding Phase-II dataset can be used for further studies and to compare predictions from various hadronic interaction models regarding the neutron content of air showers.

\begin{multicols}{2}

\end{multicols}

\newpage

\section*{The Pierre Auger Collaboration}

{\footnotesize\setlength{\baselineskip}{10pt}
\noindent
\begin{wrapfigure}[11]{l}{0.12\linewidth}
\vspace{-4pt}
\includegraphics[width=0.98\linewidth]{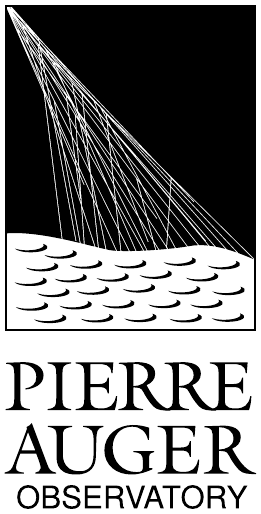}
\end{wrapfigure}
\begin{sloppypar}\noindent
A.~Abdul Halim$^{13}$,
P.~Abreu$^{70}$,
M.~Aglietta$^{53,51}$,
I.~Allekotte$^{1}$,
K.~Almeida Cheminant$^{78,77}$,
A.~Almela$^{7,12}$,
R.~Aloisio$^{44,45}$,
J.~Alvarez-Mu\~niz$^{76}$,
A.~Ambrosone$^{44}$,
J.~Ammerman Yebra$^{76}$
% ...

\end{sloppypar}
\begin{center}
\end{center}

\vspace{1ex}
\begin{description}[labelsep=0.2em,align=right,labelwidth=0.7em,labelindent=0em,leftmargin=2em,noitemsep,before={\renewcommand\makelabel[1]{##1 }}]
\item[$^{1}$] Centro At\'omico Bariloche and Instituto Balseiro (CNEA-UNCuyo-CONICET), San Carlos de Bariloche, Argentina
\item[$^{2}$] Departamento de F\'\i{}sica and Departamento de Ciencias de la Atm\'osfera y los Oc\'eanos, FCEyN, Universidad de Buenos Aires and CONICET, Buenos Aires, Argentina
\item[$^{3}$] IFLP, Universidad Nacional de La Plata and CONICET, La Plata, Argentina
\item[$^{4}$] Instituto de Astronom\'\i{}a y F\'\i{}sica del Espacio (IAFE, CONICET-UBA), Buenos Aires, Argentina
\item[$^{5}$] Instituto de F\'\i{}sica de Rosario (IFIR) -- CONICET/U.N.R.\ and Facultad de Ciencias Bioqu\'\i{}micas y Farmac\'euticas U.N.R., Rosario, Argentina
\item[$^{6}$] Instituto de Tecnolog\'\i{}as en Detecci\'on y Astropart\'\i{}culas (CNEA, CONICET, UNSAM), and Universidad Tecnol\'ogica Nacional -- Facultad Regional Mendoza (CONICET/CNEA), Mendoza, Argentina
\item[$^{7}$] Instituto de Tecnolog\'\i{}as en Detecci\'on y Astropart\'\i{}culas (CNEA, CONICET, UNSAM), Buenos Aires, Argentina
\item[$^{8}$] International Center of Advanced Studies and Instituto de Ciencias F\'\i{}sicas, ECyT-UNSAM and CONICET, Campus Miguelete -- San Mart\'\i{}n, Buenos Aires, Argentina
\item[$^{9}$] Laboratorio Atm\'osfera -- Departamento de Investigaciones en L\'aseres y sus Aplicaciones -- UNIDEF (CITEDEF-CONICET), Argentina
\item[$^{10}$] Observatorio Pierre Auger, Malarg\"ue, Argentina
\item[$^{11}$] Observatorio Pierre Auger and Comisi\'on Nacional de Energ\'\i{}a At\'omica, Malarg\"ue, Argentina
\item[$^{12}$] Universidad Tecnol\'ogica Nacional -- Facultad Regional Buenos Aires, Buenos Aires, Argentina
\item[$^{13}$] University of Adelaide, Adelaide, S.A., Australia
\item[$^{14}$] Universit\'e Libre de Bruxelles (ULB), Brussels, Belgium
\item[$^{15}$] Vrije Universiteit Brussels, Brussels, Belgium
\item[$^{16}$] Centro Brasileiro de Pesquisas Fisicas, Rio de Janeiro, RJ, Brazil
\item[$^{17}$] Centro Federal de Educa\c{c}\~ao Tecnol\'ogica Celso Suckow da Fonseca, Petropolis, Brazil
\item[$^{18}$] Instituto Federal de Educa\c{c}\~ao, Ci\^encia e Tecnologia do Rio de Janeiro (IFRJ), Brazil
\item[$^{19}$] Universidade de S\~ao Paulo, Escola de Engenharia de Lorena, Lorena, SP, Brazil
\item[$^{20}$] Universidade de S\~ao Paulo, Instituto de F\'\i{}sica de S\~ao Carlos, S\~ao Carlos, SP, Brazil
\item[$^{21}$] Universidade de S\~ao Paulo, Instituto de F\'\i{}sica, S\~ao Paulo, SP, Brazil
\item[$^{22}$] Universidade Estadual de Campinas (UNICAMP), IFGW, Campinas, SP, Brazil
\item[$^{23}$] Universidade Estadual de Feira de Santana, Feira de Santana, Brazil
\item[$^{24}$] Universidade Federal de Campina Grande, Centro de Ciencias e Tecnologia, Campina Grande, Brazil
\item[$^{25}$] Universidade Federal do ABC, Santo Andr\'e, SP, Brazil
\item[$^{26}$] Universidade Federal do Paran\'a, Setor Palotina, Palotina, Brazil
\item[$^{27}$] Universidade Federal do Rio de Janeiro, Instituto de F\'\i{}sica, Rio de Janeiro, RJ, Brazil
\item[$^{28}$] Universidad de Medell\'\i{}n, Medell\'\i{}n, Colombia
\item[$^{29}$] Universidad Industrial de Santander, Bucaramanga, Colombia
\item[$^{30}$] Charles University, Faculty of Mathematics and Physics, Institute of Particle and Nuclear Physics, Prague, Czech Republic
\item[$^{31}$] Institute of Physics of the Czech Academy of Sciences, Prague, Czech Republic
\item[$^{32}$] Palacky University, Olomouc, Czech Republic
\item[$^{33}$] CNRS/IN2P3, IJCLab, Universit\'e Paris-Saclay, Orsay, France
\item[$^{34}$] Laboratoire de Physique Nucl\'eaire et de Hautes Energies (LPNHE), Sorbonne Universit\'e, Universit\'e de Paris, CNRS-IN2P3, Paris, France
\item[$^{35}$] Univ.\ Grenoble Alpes, CNRS, Grenoble Institute of Engineering Univ.\ Grenoble Alpes, LPSC-IN2P3, 38000 Grenoble, France
\item[$^{36}$] Universit\'e Paris-Saclay, CNRS/IN2P3, IJCLab, Orsay, France
\item[$^{37}$] Bergische Universit\"at Wuppertal, Department of Physics, Wuppertal, Germany
\item[$^{38}$] Karlsruhe Institute of Technology (KIT), Institute for Experimental Particle Physics, Karlsruhe, Germany
\item[$^{39}$] Karlsruhe Institute of Technology (KIT), Institut f\"ur Prozessdatenverarbeitung und Elektronik, Karlsruhe, Germany
\item[$^{40}$] Karlsruhe Institute of Technology (KIT), Institute for Astroparticle Physics, Karlsruhe, Germany
\item[$^{41}$] RWTH Aachen University, III.\ Physikalisches Institut A, Aachen, Germany
\item[$^{42}$] Universit\"at Hamburg, II.\ Institut f\"ur Theoretische Physik, Hamburg, Germany
\item[$^{43}$] Universit\"at Siegen, Department Physik -- Experimentelle Teilchenphysik, Siegen, Germany
\item[$^{44}$] Gran Sasso Science Institute, L'Aquila, Italy
\item[$^{45}$] INFN Laboratori Nazionali del Gran Sasso, Assergi (L'Aquila), Italy
\item[$^{46}$] INFN, Sezione di Catania, Catania, Italy
\item[$^{47}$] INFN, Sezione di Lecce, Lecce, Italy
\item[$^{48}$] INFN, Sezione di Milano, Milano, Italy
\item[$^{49}$] INFN, Sezione di Napoli, Napoli, Italy
\item[$^{50}$] INFN, Sezione di Roma ``Tor Vergata'', Roma, Italy
\item[$^{51}$] INFN, Sezione di Torino, Torino, Italy
\item[$^{52}$] Istituto di Astrofisica Spaziale e Fisica Cosmica di Palermo (INAF), Palermo, Italy
\item[$^{53}$] Osservatorio Astrofisico di Torino (INAF), Torino, Italy
\item[$^{54}$] Politecnico di Milano, Dipartimento di Scienze e Tecnologie Aerospaziali , Milano, Italy
\item[$^{55}$] Universit\`a del Salento, Dipartimento di Matematica e Fisica ``E.\ De Giorgi'', Lecce, Italy
\item[$^{56}$] Universit\`a dell'Aquila, Dipartimento di Scienze Fisiche e Chimiche, L'Aquila, Italy
\item[$^{57}$] Universit\`a di Catania, Dipartimento di Fisica e Astronomia ``Ettore Majorana``, Catania, Italy
\item[$^{58}$] Universit\`a di Milano, Dipartimento di Fisica, Milano, Italy
\item[$^{59}$] Universit\`a di Napoli ``Federico II'', Dipartimento di Fisica ``Ettore Pancini'', Napoli, Italy
\item[$^{60}$] Universit\`a di Palermo, Dipartimento di Fisica e Chimica ''E.\ Segr\`e'', Palermo, Italy
\item[$^{61}$] Universit\`a di Roma ``Tor Vergata'', Dipartimento di Fisica, Roma, Italy
\item[$^{62}$] Universit\`a Torino, Dipartimento di Fisica, Torino, Italy
\item[$^{63}$] Benem\'erita Universidad Aut\'onoma de Puebla, Puebla, M\'exico
\item[$^{64}$] Unidad Profesional Interdisciplinaria en Ingenier\'\i{}a y Tecnolog\'\i{}as Avanzadas del Instituto Polit\'ecnico Nacional (UPIITA-IPN), M\'exico, D.F., M\'exico
\item[$^{65}$] Universidad Aut\'onoma de Chiapas, Tuxtla Guti\'errez, Chiapas, M\'exico
\item[$^{66}$] Universidad Michoacana de San Nicol\'as de Hidalgo, Morelia, Michoac\'an, M\'exico
\item[$^{67}$] Universidad Nacional Aut\'onoma de M\'exico, M\'exico, D.F., M\'exico
\item[$^{68}$] Institute of Nuclear Physics PAN, Krakow, Poland
\item[$^{69}$] University of \L{}\'od\'z, Faculty of High-Energy Astrophysics,\L{}\'od\'z, Poland
\item[$^{70}$] Laborat\'orio de Instrumenta\c{c}\~ao e F\'\i{}sica Experimental de Part\'\i{}culas -- LIP and Instituto Superior T\'ecnico -- IST, Universidade de Lisboa -- UL, Lisboa, Portugal
\item[$^{71}$] ``Horia Hulubei'' National Institute for Physics and Nuclear Engineering, Bucharest-Magurele, Romania
\item[$^{72}$] Institute of Space Science, Bucharest-Magurele, Romania
\item[$^{73}$] Center for Astrophysics and Cosmology (CAC), University of Nova Gorica, Nova Gorica, Slovenia
\item[$^{74}$] Experimental Particle Physics Department, J.\ Stefan Institute, Ljubljana, Slovenia
\item[$^{75}$] Universidad de Granada and C.A.F.P.E., Granada, Spain
\item[$^{76}$] Instituto Galego de F\'\i{}sica de Altas Enerx\'\i{}as (IGFAE), Universidade de Santiago de Compostela, Santiago de Compostela, Spain
\item[$^{77}$] IMAPP, Radboud University Nijmegen, Nijmegen, The Netherlands
\item[$^{78}$] Nationaal Instituut voor Kernfysica en Hoge Energie Fysica (NIKHEF), Science Park, Amsterdam, The Netherlands
\item[$^{79}$] Stichting Astronomisch Onderzoek in Nederland (ASTRON), Dwingeloo, The Netherlands
\item[$^{80}$] Universiteit van Amsterdam, Faculty of Science, Amsterdam, The Netherlands
\item[$^{81}$] Case Western Reserve University, Cleveland, OH, USA
\item[$^{82}$] Colorado School of Mines, Golden, CO, USA
\item[$^{83}$] Department of Physics and Astronomy, Lehman College, City University of New York, Bronx, NY, USA
\item[$^{84}$] Michigan Technological University, Houghton, MI, USA
\item[$^{85}$] New York University, New York, NY, USA
\item[$^{86}$] University of Chicago, Enrico Fermi Institute, Chicago, IL, USA
\item[$^{87}$] University of Delaware, Department of Physics and Astronomy, Bartol Research Institute, Newark, DE, USA
\item[] -----
\item[$^{a}$] Max-Planck-Institut f\"ur Radioastronomie, Bonn, Germany
\item[$^{b}$] also at Kapteyn Institute, University of Groningen, Groningen, The Netherlands
\item[$^{c}$] School of Physics and Astronomy, University of Leeds, Leeds, United Kingdom
\item[$^{d}$] Fermi National Accelerator Laboratory, Fermilab, Batavia, IL, USA
\item[$^{e}$] Pennsylvania State University, University Park, PA, USA
\item[$^{f}$] Colorado State University, Fort Collins, CO, USA
\item[$^{g}$] Louisiana State University, Baton Rouge, LA, USA
\item[$^{h}$] now at Graduate School of Science, Osaka Metropolitan University, Osaka, Japan
\item[$^{i}$] Institut universitaire de France (IUF), France
\item[$^{j}$] now at Technische Universit\"at Dortmund and Ruhr-Universit\"at Bochum, Dortmund and Bochum, Germany
\end{description}

\section*{Acknowledgments}

\begin{sloppypar}
The successful installation, commissioning, and operation of the Pierre
Auger Observatory would not have been possible without the strong
commitment and effort from the technical and administrative staff in
Malarg\"ue. We are very grateful to the following agencies and
organizations for financial support:
\end{sloppypar}

\begin{sloppypar}
Argentina -- Comisi\'on Nacional de Energ\'\i{}a At\'omica; Agencia Nacional de
Promoci\'on Cient\'\i{}fica y Tecnol\'ogica (ANPCyT); Consejo Nacional de
Investigaciones Cient\'\i{}ficas y T\'ecnicas (CONICET); Gobierno de la
Provincia de Mendoza; Municipalidad de Malarg\"ue; NDM Holdings and Valle
Las Le\~nas; in gratitude for their continuing cooperation over land
access; Australia -- the Australian Research Council; Belgium -- Fonds
de la Recherche Scientifique (FNRS); Research Foundation Flanders (FWO),
Marie Curie Action of the European Union Grant No.~101107047; Brazil --
Conselho Nacional de Desenvolvimento Cient\'\i{}fico e Tecnol\'ogico (CNPq);
Financiadora de Estudos e Projetos (FINEP); Funda\c{c}\~ao de Amparo \`a
Pesquisa do Estado de Rio de Janeiro (FAPERJ); S\~ao Paulo Research
Foundation (FAPESP) Grants No.~2019/10151-2, No.~2010/07359-6 and
No.~1999/05404-3; Minist\'erio da Ci\^encia, Tecnologia, Inova\c{c}\~oes e
Comunica\c{c}\~oes (MCTIC); Czech Republic -- GACR 24-13049S, CAS LQ100102401,
MEYS LM2023032, CZ.02.1.01/0.0/0.0/16{\textunderscore}013/0001402,
CZ.02.1.01/0.0/0.0/18{\textunderscore}046/0016010 and
CZ.02.1.01/0.0/0.0/17{\textunderscore}049/0008422 and CZ.02.01.01/00/22{\textunderscore}008/0004632;
France -- Centre de Calcul IN2P3/CNRS; Centre National de la Recherche
Scientifique (CNRS); Conseil R\'egional Ile-de-France; D\'epartement
Physique Nucl\'eaire et Corpusculaire (PNC-IN2P3/CNRS); D\'epartement
Sciences de l'Univers (SDU-INSU/CNRS); Institut Lagrange de Paris (ILP)
Grant No.~LABEX ANR-10-LABX-63 within the Investissements d'Avenir
Programme Grant No.~ANR-11-IDEX-0004-02; Germany -- Bundesministerium
f\"ur Bildung und Forschung (BMBF); Deutsche Forschungsgemeinschaft (DFG);
Finanzministerium Baden-W\"urttemberg; Helmholtz Alliance for
Astroparticle Physics (HAP); Helmholtz-Gemeinschaft Deutscher
Forschungszentren (HGF); Ministerium f\"ur Kultur und Wissenschaft des
Landes Nordrhein-Westfalen; Ministerium f\"ur Wissenschaft, Forschung und
Kunst des Landes Baden-W\"urttemberg; Italy -- Istituto Nazionale di
Fisica Nucleare (INFN); Istituto Nazionale di Astrofisica (INAF);
Ministero dell'Universit\`a e della Ricerca (MUR); CETEMPS Center of
Excellence; Ministero degli Affari Esteri (MAE), ICSC Centro Nazionale
di Ricerca in High Performance Computing, Big Data and Quantum
Computing, funded by European Union NextGenerationEU, reference code
CN{\textunderscore}00000013; M\'exico -- Consejo Nacional de Ciencia y Tecnolog\'\i{}a
(CONACYT) No.~167733; Universidad Nacional Aut\'onoma de M\'exico (UNAM);
PAPIIT DGAPA-UNAM; The Netherlands -- Ministry of Education, Culture and
Science; Netherlands Organisation for Scientific Research (NWO); Dutch
national e-infrastructure with the support of SURF Cooperative; Poland
-- Ministry of Education and Science, grants No.~DIR/WK/2018/11 and
2022/WK/12; National Science Centre, grants No.~2016/22/M/ST9/00198,
2016/23/B/ST9/01635, 2020/39/B/ST9/01398, and 2022/45/B/ST9/02163;
Portugal -- Portuguese national funds and FEDER funds within Programa
Operacional Factores de Competitividade through Funda\c{c}\~ao para a Ci\^encia
e a Tecnologia (COMPETE); Romania -- Ministry of Research, Innovation
and Digitization, CNCS-UEFISCDI, contract no.~30N/2023 under Romanian
National Core Program LAPLAS VII, grant no.~PN 23 21 01 02 and project
number PN-III-P1-1.1-TE-2021-0924/TE57/2022, within PNCDI III; Slovenia
-- Slovenian Research Agency, grants P1-0031, P1-0385, I0-0033, N1-0111;
Spain -- Ministerio de Ciencia e Innovaci\'on/Agencia Estatal de
Investigaci\'on (PID2019-105544GB-I00, PID2022-140510NB-I00 and
RYC2019-027017-I), Xunta de Galicia (CIGUS Network of Research Centers,
Consolidaci\'on 2021 GRC GI-2033, ED431C-2021/22 and ED431F-2022/15),
Junta de Andaluc\'\i{}a (SOMM17/6104/UGR and P18-FR-4314), and the European
Union (Marie Sklodowska-Curie 101065027 and ERDF); USA -- Department of
Energy, Contracts No.~DE-AC02-07CH11359, No.~DE-FR02-04ER41300,
No.~DE-FG02-99ER41107 and No.~DE-SC0011689; National Science Foundation,
Grant No.~0450696, and NSF-2013199; The Grainger Foundation; Marie
Curie-IRSES/EPLANET; European Particle Physics Latin American Network;
and UNESCO.
\end{sloppypar}

}

\end{document}